\documentclass[sigconf,nonacm]{acmart}
\pdfoutput=1

\setcopyright{acmcopyright}
\copyrightyear{2022}
\acmYear{2022}
\acmDOI{XXXXXXX.XXXXXXX}

\acmConference[FAccTRec '22]{5th FAccTRec Workshop on Responsible Recommendation}{September 2022}{Seattle, WA}
\acmBooktitle{Woodstock '18: ACM Symposium on Neural Gaze Detection,
 June 03--05, 2018, Woodstock, NY} 

\usepackage{xspace}
\newcommand{\RS}{\texttt{RS}\xspace}

\begin{document}

\title{Matching Consumer Fairness Objectives \& Strategies for RecSys}
\titlenote{Presented at FAccTRec 2022.}
\fancyhead[LE,RO]{\small\sffamily FAccTRec 2022}

\author{Michael D. Ekstrand}
\authornote{Both authors contributed equally to this paper.}
\email{ekstrand@acm.org}
\orcid{0000-0003-2467-0108}
\affiliation{%
  \institution{People and Information Research Team \\ Boise State University}
  \streetaddress{1910 University Drive}
  \city{Boise}
  \state{Idaho}
  \country{USA}
  \postcode{83725-2055}
}

\author{Maria Soledad Pera}
\authornotemark[2]
\authornote{Work begun while at Boise State University.}
\email{M.S.Pera@TUDelft.nl}
\orcid{0000-0002-2008-9204}
\affiliation{%
  \institution{Web Information Systems Group \\ Delft University of Technology}
  \city{Delft}
  \country{The Netherlands}
}

\renewcommand{\shortauthors}{Ekstrand and Pera}

\begin{abstract}
The last several years have brought a growing body of work on ensuring that recommender systems are in some sense \textit{consumer-fair} --- that is, they provide comparable quality of service, accuracy of representation, and other effects to their users.
However, there are many different strategies to make systems more fair and a range of intervention points.
In this position paper, we build on ongoing work to highlight the need for researchers and practitioners to attend to the details of their application, users, and the fairness objective they aim to achieve, and adopt interventions that are appropriate to the situation.
We argue that consumer fairness should be a creative endeavor flowing from the particularities of the specific problem to be solved.
\end{abstract}

\maketitle

\section{Paths to Consumer Fairness}

Fair recommendation is a complex and multi-sided problem \citep{Sonboli2022-cx, fnt-fairness}, with a significant focus on providing a fair experience to one or both of two main stakeholders: producers (who provide the items or services to be suggested) and users (who consume the provided recommendations) \cite{Burke2017-ne}. We are particularly interested in the latter group, for whom recommender systems (\textbf{\RS}) have to offer appealing items while considering that ``the best items
for one user may be different than those for another'' \cite{Burke2017-ne}.
\textit{Consumer fairness} \citep{Boratto2022-pe} is the aspect of fairness concerned with ensuring that the users (or ``consumers'') of a \RS\ are treated fairly in the quantitative and/or qualitative aspects of their experience.
The relevant literature considers several ideas of what it means to be ``fair'' to consumers, along with different techniques to measure or attain such fairness; one particularly common goal is to ensure that certain users or groups of users do not receive a systematically lower-quality or less-useful experience than others \citep{ekstrand-cool-kids, Mehrotra2017-ns, Neophytou2022-cl}.

This interest mirrors a line of work on specific user audiences. \citet{ekstrand-cool-kids} show recommender performance can differ between users of different genders and ages. Explorations of children's media use \citep{Spear2021-lr, Milton2020-ce} reveal that preferred traits in songs and books vary from childhood to early adulthood, indirectly urging \RS work to treat ``children'' not as a monolithic entity, but as individuals to better serve them. Researchers have suggested going beyond traditional popularity- or collaborative-filtering algorithms that would inevitably prioritize the majority of the consumers (i.e., adults) to explicitly consider factors like the readability levels (comprehension), familiarity with concepts covered in the classroom (learning), and explainability (engagement and improve task performance), if suggestions are to be suitable--and therefore apt for consumption \cite{Murgia2019-ez, Milton2019-lh, Pera2014-ww, Tsiakas2020-hm,Pera2019-fh}. Literature bringing awareness to autism \cite{Mauro2020-be,Ng2018-lm, Banskota2020-vh} emphasizes that \RS\ should account for user-specific sensory aversions or skill limitations of recommended items are to be compatible with what these users require, and hence useful.

Several concerns from the broader \RS literature can also be regarded as forms of consumer fairness. Examples include macro-averaging evaluation metrics by user \citep{Gunawardana2022-aw, Tague-Sutcliffe1992-bi} to assess the experience of all users instead of emphasizing highly-active users \citep{Ekstrand2010-wg, fnt-fairness} and providing good results to new users \citep{fnt-fairness}.

The works presented thus far share the common goal of providing effective, often personalized, experiences to all their users. They do so through a variety of definitions, methods, and points of intervention (where the \RS is changed to advance the goal).
\citet[\S 5]{fnt-fairness} have cataloged many of the existing strategies and noted some challenges in matching a strategy to specific fairness objectives.
Expanding on that argument, our proposition in this paper is that \textbf{researchers and practitioners need to select interventions that are appropriate to the specific fairness goal(s) and particularities of an application context}.
More importantly, we hope to see a robust discussion between researchers, practitioners, and stakeholder representatives from different disciplinary perspectives to understand how best to promote \RS\ that are ``good'' --- in multiple relevant ways --- for everyone who uses them.

\section{Types of Fairness Objectives}
\label{sec:goals}

Numerous fairness objectives have been studied under the banner of consumer fairness.
Perhaps the most well-known is \textit{equity of utility}: ensuring that a \RS\ (or other information access system) provides comparable quality of service to all users or groups of users \citep[e.g.][]{Mehrotra2017-ns,ekstrand-cool-kids, Li2021-eq, Spear2021-lr,Huibers2019-el}, typically measured by online or offline effective measures such as nDCG or click-through rate. A related objective is \textit{equity of usability}: ensuring that people can actually use the system, either in addition to or independent of considering equity in the utility of results \citep{Anuyah2019-rv, Ng2018-lm, Mauro2020-be,Huibers2021-fo,Rothschild2019-xc}.
Accessibility is a clear concern here, as a system that does not work with screen readers, for example, cannot be used as easily by visually-impaired users \cite{Vtyurina2019-hz, Fourney2018-vx, Milton2021-rp, Andrunyk2019-mk, Kreiss2022-lf, Chen2008-ys}.
Other works focus on attending to \textit{specific information needs} that a particular group of users may have that are not effectively met by systems more attuned to needs common among the majority of the population \cite{Rothschild2019-xc,Azpiazu2017-ue,Andrunyk2020-information}.

Looking past the effectiveness and usability of a \RS, some consumer fairness work has looked at issues of \textit{fair representation} or \textit{representational harms} \citep{Crawford2017-js}, in terms of either the \RS's internal representation of the user (e.g. avoiding user embeddings that may lead to stereotyped recommendations \citep{Beutel2017-lx}) or the recommended items themselves.
One example of this last concern is the objective of \textit{recommendation independence} \citep{Kamishima2018-nn}: this goal is satisfied if the probability of a particular item being recommended to the user is independent of their gender or other protected status.

Effective consumer fairness must begin by identifying an objective to pursue or problem to solve, as the choice of operationalization and intervention (\S \ref{sec:points}) follows from the objective \citep{Ekstrand2022-handbook-chapter}.

\section{Intervention Strategies}
\label{sec:points}

Prior work has proposed various strategies to advance one or more fairness objectives (\S \ref{sec:goals}). Here, we mention some salient ones, grouped by the stage of the \RS at which they intervene.

\paragraph{Design interventions}
Designing \RS to adapt to users in the quest for consumer fairness involves the use of multiple interfaces that are matched to users' needs. For example, \citet{Deldjoo2017-qs} proposed a child-oriented TV/movie recommendation interface for in-home set-top boxes that incorporated tangible interaction: the child could hold up a toy truck to get recommendations for shows about trucks. 
Another common alternative is to detect the particular group a user belongs to and adapt \RS\ behavior and/or interface to the corresponding group. Practical applications of this strategy include, upon identification of the grade or skill of the target user, modifying the types of queries that are recommended \citet{Madrazo_Azpiazu2018-ld}, showcasing different multi-modal cues to point users towards suitable spelling suggestions \cite{Downs2021-ul}, or adapting presented choices to enable knowledge acquisition \cite{Rothschild2019-xc}. 

\paragraph{Algorithmic interventions}
\begin{sloppypar}
Modifying recommendation algorithms is common. This means, for instance, including the inter-user equity objective into the loss function \citep{Wu2022-fi, Huang2020-na, Wang2021-zi}, sometimes through a regularization term \citep{Kamishima2018-nn, Yao2017-vz}.
Reranking \citep{Li2021-eq} can also reduce gaps in utility by post-processing recommendations from an existing model.
These can be applied to many objectives beyond equity of utility. Penalizing dependence between recommended items and user attributes \citep{Kamishima2018-nn} is another alternative.
\end{sloppypar}
Adversarial learning methods can also help reduce unfairness. \citet{Beutel2017-lx} use a discriminator to learn user embeddings that are not predictive of sensitive attributes such as race or gender; more broadly, fair representation learning \citep{Madras2018-at, Zemel2013-kq} can be applied to consumer fairness \citep{Wu2021-on}.
There are also many other algorithmic strategies considered as well, such as changing neighborhoods \citep{Burke2018-fm}.

\paragraph{Data interventions}

Some strategies manipulate the \RS's input data to improve fairness, e.g. by injecting fake user profiles \citep{Rastegarpanah2019-qd} or removing spam reviews \citep{Shrestha2021-hq}.

\paragraph{Process interventions}

Improvements to engineering and quality assurance processes can be useful for providing consumer fairness.
Regular auditing for violations of fairness objectives \citep{Holstein2019-ho}, through disaggregated evaluations \citep{Mehrotra2017-ns, Barocas2021-tv} or other means, identify problems and help detect regressions on past fairness improvements.

The engineering process is another place to improve a system's fairness. Little has been little published on this, but studies that reveal \emph{why} a fairness problem occurs may enable engineers and model owners to identify and prioritize software improvements that will address the problem, even if they are not directly fairness interventions.
For example, if a music recommender performs poorly for users from a particular region due to lower-quality song metadata, investing in that data could improve equity of utility.

\paragraph{Marketplace interventions}
Consumer fairness can also call for the development of new \RS targeted at under-served groups. This can be done either by new entrants to the market or existing firms seeking to shore up their market position. Consider popular sites like Goodreads and Amazon: the segment of their user base producing the most interactions, and hence driving recommendation algorithms, are adults. In turn, the resulting experience may not suit children. Some startups are trying to fill this perceived gap by creating new sites specifically for children; examples include ABC Mouse \cite{abcmouse}, BiblioNasium \cite{biblionasium}, or Pickatale \cite{pickatale}. As for examples of sites aiming to expand their target audiences, we find Netflix offering recommendations specific to children and families \cite{netflixKids} or Spotify, which now offers Spotify Kids \cite{spotifyKids} as an alternative to better support children.

\section{Matching Objectives and Strategies}

Our central proposition in this paper is that the choice of \textit{where} in the \RS and its sociotechnical context to intervene, and \textit{how} to intervene at that point, needs to be well-matched to the specific fairness objective and details of the application, domain, and users.

Some pairings of strategies and outcomes are better-matched than others.
E.g., auditing differences in utility \citep{ekstrand-cool-kids, Mehrotra2017-ns, Barocas2021-tv} can identify unfair utility and  provide an empirical starting point for many potential strategies, including design and process interventions, but not every intervention strategy is likely a good fit for this objective.  \citet{fnt-fairness} note that useful recommendations in most domains are not a subtractable good \citep{Becker1995-zo} (users do not compete with each other for good recommendations). The inequity itself is not the problem, but rather a symptom of the system not providing some of its users with good recommendations. Training to minimize differences in utility \citep[e.g.][]{Li2021-eq, Huang2020-na, Naghiaei2022-ct} can ensure equity, but at the risk of placing users in competition with each other, sometimes with significant majority-group utility loss \citep{Li2021-eq}.
Positive-sum rather than zero-sum utility aggregates avoids the competition problem \citep{Wang2021-zi}, as do interventions that seek to directly address the causes of under-serving a segment of the user base \citep{fnt-fairness}. 

A better-matched pairing involving algorithmic intervention is \citet{Beutel2017-lx}'s use of adversarial learning to remove unwanted correlations between user embeddings and user group membership in hopes of producing less stereotypical recommendations. 

A single objective may have significantly more complexity and nuance than is accounted for by simple strategies.
For example, what counts as a good recommendation may differ between groups \citep{Huibers2021-fo} and contexts. Here, pursuing an objective such as equity of utility should consider whether metrics accurately measure utility across the varied constituencies and contexts in an \RS's usage.

We invite the broad community of people concerned with ensuring fair access to information through \RS and related information access systems to think carefully and interdisciplinarily about the specific problems to be solved and select appropriate, not just convenient or familiar, interventions.
Further research is needed to understand how to implement the various interventions in \S\ref{sec:points} (and more not listed) most effectively, and to more thoroughly decompose the problem space of consumer fairness.
Such research will identify when different interventions may or may not be appropriate, and provide evidence-based guidance for future practice.

\begin{acks}
Partially supported by National Science Foundation grant 17-51278.
\end{acks}
\balance

\bibliographystyle{ACM-Reference-Format}
\bibliography{References-Paperpile-Clean,url-refs}


\begin{thebibliography}{59}


\ifx \showCODEN    \undefined \def \showCODEN     #1{\unskip}     \fi
\ifx \showDOI      \undefined \def \showDOI       #1{#1}\fi
\ifx \showISBNx    \undefined \def \showISBNx     #1{\unskip}     \fi
\ifx \showISBNxiii \undefined \def \showISBNxiii  #1{\unskip}     \fi
\ifx \showISSN     \undefined \def \showISSN      #1{\unskip}     \fi
\ifx \showLCCN     \undefined \def \showLCCN      #1{\unskip}     \fi
\ifx \shownote     \undefined \def \shownote      #1{#1}          \fi
\ifx \showarticletitle \undefined \def \showarticletitle #1{#1}   \fi
\ifx \showURL      \undefined \def \showURL       {\relax}        \fi
\providecommand\bibfield[2]{#2}
\providecommand\bibinfo[2]{#2}
\providecommand\natexlab[1]{#1}
\providecommand\showeprint[2][]{arXiv:#2}

\bibitem[abc(2022)]%
        {abcmouse}
 \bibinfo{year}{2022}\natexlab{}.
\newblock \bibinfo{title}{ABC Mouse}.
\newblock \bibinfo{howpublished}{\url{https://www.abcmouse.com/}}.
\newblock


\bibitem[bib(2022)]%
        {biblionasium}
 \bibinfo{year}{2022}\natexlab{}.
\newblock \bibinfo{title}{BiblioNasium}.
\newblock \bibinfo{howpublished}{\url{https://www.biblionasium.com/}}.
\newblock


\bibitem[pic(2022)]%
        {pickatale}
 \bibinfo{year}{2022}\natexlab{}.
\newblock \bibinfo{title}{Pickatale}.
\newblock \bibinfo{howpublished}{\url{https://pickatale.com/}}.
\newblock


\bibitem[Andrunyk et~al\mbox{.}(2019)]%
        {Andrunyk2019-mk}
\bibfield{author}{\bibinfo{person}{Vasyl Andrunyk}, \bibinfo{person}{Volodymyr
  Pasichnyk}, \bibinfo{person}{Tetiana Shestakevych}, {and}
  \bibinfo{person}{Natalya Antonyuk}.} \bibinfo{year}{2019}\natexlab{}.
\newblock \showarticletitle{Modeling the Recommender System for the Synthesis
  of Information and Technology Complexes for the Education of Students with
  Autism}. In \bibinfo{booktitle}{\emph{Proceedings of {IEEE} 14th
  International Conference on Computer Sciences and Information Technologies
  ({CSIT})}}, Vol.~\bibinfo{volume}{3}. \bibinfo{pages}{183--186}.
\newblock
\urldef\tempurl%
\url{https://doi.org/10.1109/STC-CSIT.2019.8929776}
\showDOI{\tempurl}


\bibitem[Andrunyk et~al\mbox{.}(2020)]%
        {Andrunyk2020-information}
\bibfield{author}{\bibinfo{person}{Vasyl Andrunyk}, \bibinfo{person}{Tetiana
  Shestakevych}, \bibinfo{person}{Volodymyr Pasichnyk}, {and}
  \bibinfo{person}{Nataliia Kunanets}.} \bibinfo{year}{2020}\natexlab{}.
\newblock \showarticletitle{Information Technologies for Teaching Children with
  {ASD}}. In \bibinfo{booktitle}{\emph{Advances in Computer Science for
  Engineering and Education {II}}}. \bibinfo{publisher}{Springer International
  Publishing}, \bibinfo{pages}{523--533}.
\newblock
\urldef\tempurl%
\url{https://doi.org/10.1007/978-3-030-16621-2\_49}
\showDOI{\tempurl}


\bibitem[Anuyah et~al\mbox{.}(2019)]%
        {Anuyah2019-rv}
\bibfield{author}{\bibinfo{person}{Oghenemaro Anuyah}, \bibinfo{person}{Michael
  Green}, \bibinfo{person}{Ashlee Milton}, {and} \bibinfo{person}{Maria~Soledad
  Pera}.} \bibinfo{year}{2019}\natexlab{}.
\newblock \showarticletitle{The Need for a Comprehensive Strategy to Evaluate
  Search Engine Performance in the Classroom}. In
  \bibinfo{booktitle}{\emph{{KidRec} '19: Workshop in International and
  Interdisciplinary Perspectives on Children \& Recommender and Information
  Retrieval Systems, Co-located with {ACM} {IDC}}}.
\newblock
\urldef\tempurl%
\url{https://kidrec.github.io/papers/KidRec_2019_paper_1.pdf}
\showURL{%
\tempurl}


\bibitem[Azpiazu et~al\mbox{.}(2017)]%
        {Azpiazu2017-ue}
\bibfield{author}{\bibinfo{person}{Ion~Madrazo Azpiazu},
  \bibinfo{person}{Nevena Dragovic}, \bibinfo{person}{Maria~Soledad Pera},
  {and} \bibinfo{person}{Jerry~Alan Fails}.} \bibinfo{year}{2017}\natexlab{}.
\newblock \showarticletitle{Online searching and learning: {YUM} and other
  search tools for children and teachers}.
\newblock \bibinfo{journal}{\emph{Information Retrieval Journal}}
  \bibinfo{volume}{20}, \bibinfo{number}{5} (\bibinfo{year}{2017}),
  \bibinfo{pages}{524--545}.
\newblock
\showISSN{1386-4564, 1573-7659}
\urldef\tempurl%
\url{https://doi.org/10.1007/s10791-017-9310-1}
\showDOI{\tempurl}


\bibitem[Banskota and Ng(2020)]%
        {Banskota2020-vh}
\bibfield{author}{\bibinfo{person}{Alisha Banskota} {and}
  \bibinfo{person}{Yiu-Kai Ng}.} \bibinfo{year}{2020}\natexlab{}.
\newblock \showarticletitle{Recommending Video Games to Adults with Autism
  Spectrum Disorder for {Social-Skill} Enhancement}. In
  \bibinfo{booktitle}{\emph{Proceedings of the 28th {ACM} Conference on User
  Modeling, Adaptation and Personalization}} \emph{(\bibinfo{series}{UMAP
  '20})}. \bibinfo{publisher}{ACM}, \bibinfo{pages}{14--22}.
\newblock
\showISBNx{9781450368612}
\urldef\tempurl%
\url{https://doi.org/10.1145/3340631.3394867}
\showDOI{\tempurl}


\bibitem[Barocas et~al\mbox{.}(2021)]%
        {Barocas2021-tv}
\bibfield{author}{\bibinfo{person}{Solon Barocas}, \bibinfo{person}{Anhong
  Guo}, \bibinfo{person}{Ece Kamar}, \bibinfo{person}{Jacquelyn Krones},
  \bibinfo{person}{Meredith~Ringel Morris}, \bibinfo{person}{Jennifer~Wortman
  Vaughan}, \bibinfo{person}{W~Duncan Wadsworth}, {and} \bibinfo{person}{Hanna
  Wallach}.} \bibinfo{year}{2021}\natexlab{}.
\newblock \showarticletitle{Designing Disaggregated Evaluations of {AI}
  Systems: Choices, Considerations, and Tradeoffs}. In
  \bibinfo{booktitle}{\emph{Proceedings of the 2021 {AAAI/ACM} Conference on
  {AI}, Ethics, and Society}} \emph{(\bibinfo{series}{AIES '21})}.
  \bibinfo{publisher}{ACM}, \bibinfo{pages}{368--378}.
\newblock
\showISBNx{9781450384735}
\urldef\tempurl%
\url{https://doi.org/10.1145/3461702.3462610}
\showDOI{\tempurl}


\bibitem[Becker and Ostrom(1995)]%
        {Becker1995-zo}
\bibfield{author}{\bibinfo{person}{C~Dustin Becker} {and}
  \bibinfo{person}{Elinor Ostrom}.} \bibinfo{year}{1995}\natexlab{}.
\newblock \showarticletitle{Human Ecology and Resource Sustainability: The
  Importance of Institutional Diversity}.
\newblock \bibinfo{journal}{\emph{Annual Review of Ecology and Systematics}}
  \bibinfo{volume}{26}, \bibinfo{number}{1} (\bibinfo{year}{1995}),
  \bibinfo{pages}{113--133}.
\newblock
\showISSN{0066-4162}
\urldef\tempurl%
\url{https://doi.org/10.1146/annurev.es.26.110195.000553}
\showDOI{\tempurl}


\bibitem[Beutel et~al\mbox{.}(2017)]%
        {Beutel2017-lx}
\bibfield{author}{\bibinfo{person}{Alex Beutel}, \bibinfo{person}{Jilin Chen},
  \bibinfo{person}{Zhe Zhao}, {and} \bibinfo{person}{Ed~H Chi}.}
  \bibinfo{year}{2017}\natexlab{}.
\newblock \showarticletitle{Data Decisions and Theoretical Implications when
  Adversarially Learning Fair Representations}.
\newblock  (\bibinfo{year}{2017}).
\newblock
\urldef\tempurl%
\url{http://arxiv.org/abs/1707.00075}
\showURL{%
\tempurl}


\bibitem[Boratto et~al\mbox{.}(2022)]%
        {Boratto2022-pe}
\bibfield{author}{\bibinfo{person}{Ludovico Boratto}, \bibinfo{person}{Gianni
  Fenu}, \bibinfo{person}{Mirko Marras}, {and} \bibinfo{person}{Giacomo
  Medda}.} \bibinfo{year}{2022}\natexlab{}.
\newblock \showarticletitle{Consumer Fairness in Recommender Systems:
  Contextualizing Definitions and Mitigations}.
\newblock In \bibinfo{booktitle}{\emph{Lecture Notes in Computer Science}}.
  \bibinfo{publisher}{Springer International Publishing},
  \bibinfo{pages}{552--566}.
\newblock
\showISBNx{9783030997359, 9783030997366}
\showISSN{0302-9743, 1611-3349}
\urldef\tempurl%
\url{https://doi.org/10.1007/978-3-030-99736-6\_37}
\showDOI{\tempurl}


\bibitem[Burke(2017)]%
        {Burke2017-ne}
\bibfield{author}{\bibinfo{person}{Robin Burke}.}
  \bibinfo{year}{2017}\natexlab{}.
\newblock \showarticletitle{Multisided Fairness for Recommendation}.
\newblock  (\bibinfo{year}{2017}).
\newblock
\urldef\tempurl%
\url{http://arxiv.org/abs/1707.00093}
\showURL{%
\tempurl}


\bibitem[Burke et~al\mbox{.}(2018)]%
        {Burke2018-fm}
\bibfield{author}{\bibinfo{person}{Robin Burke}, \bibinfo{person}{Nasim
  Sonboli}, {and} \bibinfo{person}{Aldo Ordonez-Gauger}.}
  \bibinfo{year}{2018}\natexlab{}.
\newblock \showarticletitle{Balanced Neighborhoods for Multi-sided Fairness in
  Recommendation}. In \bibinfo{booktitle}{\emph{Proceedings of the 1st
  Conference on Fairness, Accountability and Transparency}}
  \emph{(\bibinfo{series}{Proceedings of Machine Learning Research},
  Vol.~\bibinfo{volume}{81})}, \bibfield{editor}{\bibinfo{person}{Sorelle~A
  Friedler} {and} \bibinfo{person}{Christo Wilson}} (Eds.).
  \bibinfo{publisher}{PMLR}, \bibinfo{pages}{202--214}.
\newblock
\urldef\tempurl%
\url{http://proceedings.mlr.press/v81/burke18a.html}
\showURL{%
\tempurl}


\bibitem[Chen et~al\mbox{.}(2008)]%
        {Chen2008-ys}
\bibfield{author}{\bibinfo{person}{Wei Chen}, \bibinfo{person}{Li-Jun Zhang},
  \bibinfo{person}{Can Wang}, \bibinfo{person}{Chun Chen}, {and}
  \bibinfo{person}{Jia-Jun Bu}.} \bibinfo{year}{2008}\natexlab{}.
\newblock \showarticletitle{Pervasive Web News Recommendation for Visually
  Impaired People}. In \bibinfo{booktitle}{\emph{Proceedings of {IEEE/WIC/ACM}
  International Conference on Web Intelligence and Intelligent Agent
  Technology}} \emph{(\bibinfo{series}{WI-IAT '08})}.
  \bibinfo{pages}{119--122}.
\newblock
\urldef\tempurl%
\url{https://doi.org/10.1109/WIIAT.2008.43}
\showDOI{\tempurl}


\bibitem[Crawford(2017)]%
        {Crawford2017-js}
\bibfield{author}{\bibinfo{person}{Kate Crawford}.}
  \bibinfo{year}{2017}\natexlab{}.
\newblock \bibinfo{title}{The Trouble with Bias}.
\newblock \bibinfo{howpublished}{Neural Information Processing Systems 2017}.
\newblock
\urldef\tempurl%
\url{https://youtu.be/fMym_BKWQzk}
\showURL{%
\tempurl}


\bibitem[Deldjoo et~al\mbox{.}(2017)]%
        {Deldjoo2017-qs}
\bibfield{author}{\bibinfo{person}{Yashar Deldjoo}, \bibinfo{person}{Cristina
  Fr{\`a}}, \bibinfo{person}{Massimo Valla}, \bibinfo{person}{Antonio
  Paladini}, \bibinfo{person}{Davide Anghileri}, \bibinfo{person}{Mustafa~Anil
  Tuncel}, \bibinfo{person}{Franca Garzotto}, {and} \bibinfo{person}{Paolo
  Cremonesi}.} \bibinfo{year}{2017}\natexlab{}.
\newblock \showarticletitle{Enhancing Children's Experience with Recommendation
  Systems}. In \bibinfo{booktitle}{\emph{{KidRec} 2017}}.
\newblock
\urldef\tempurl%
\url{https://yasdel.github.io/files/KidRec17_deldjoo.pdf}
\showURL{%
\tempurl}


\bibitem[Downs et~al\mbox{.}(2021)]%
        {Downs2021-ul}
\bibfield{author}{\bibinfo{person}{Brody Downs}, \bibinfo{person}{Maria~Soledad
  Pera}, \bibinfo{person}{Katherine~Landau Wright}, \bibinfo{person}{Casey
  Kennington}, {and} \bibinfo{person}{Jerry~Alan Fails}.}
  \bibinfo{year}{2021}\natexlab{}.
\newblock \showarticletitle{{KidSpell}: Making a difference in spellchecking
  for children}.
\newblock \bibinfo{journal}{\emph{International Journal of Child-Computer
  Interaction}}  \bibinfo{volume}{32} (\bibinfo{year}{2021}),
  \bibinfo{pages}{100373}.
\newblock
\showISSN{2212-8689}
\urldef\tempurl%
\url{https://doi.org/10.1016/j.ijcci.2021.100373}
\showDOI{\tempurl}


\bibitem[Ekstrand et~al\mbox{.}(2010)]%
        {Ekstrand2010-wg}
\bibfield{author}{\bibinfo{person}{Michael Ekstrand}, \bibinfo{person}{John
  Riedl}, {and} \bibinfo{person}{Joseph~A Konstan}.}
  \bibinfo{year}{2010}\natexlab{}.
\newblock \showarticletitle{Collaborative Filtering Recommender Systems}.
\newblock \bibinfo{journal}{\emph{Foundations and Trends\textregistered{} in
  Human-Computer Interaction}} \bibinfo{volume}{4}, \bibinfo{number}{2}
  (\bibinfo{year}{2010}), \bibinfo{pages}{81--173}.
\newblock
\showISSN{1551-3955}
\urldef\tempurl%
\url{https://doi.org/10.1561/1100000009}
\showDOI{\tempurl}


\bibitem[Ekstrand et~al\mbox{.}(2022a)]%
        {fnt-fairness}
\bibfield{author}{\bibinfo{person}{Michael~D Ekstrand},
  \bibinfo{person}{Anubrata Das}, \bibinfo{person}{Robin Burke}, {and}
  \bibinfo{person}{Fernando Diaz}.} \bibinfo{year}{2022}\natexlab{a}.
\newblock \showarticletitle{Fairness in Information Access Systems}.
\newblock \bibinfo{journal}{\emph{Foundations and Trends\textregistered{} in
  Information Retrieval}} \bibinfo{volume}{16}, \bibinfo{number}{1-2}
  (\bibinfo{year}{2022}), \bibinfo{pages}{1--177}.
\newblock
\showISSN{1554-0669}
\urldef\tempurl%
\url{https://doi.org/10.1561/1500000079}
\showDOI{\tempurl}


\bibitem[Ekstrand et~al\mbox{.}(2022b)]%
        {Ekstrand2022-handbook-chapter}
\bibfield{author}{\bibinfo{person}{Michael~D Ekstrand},
  \bibinfo{person}{Anubrata Das}, \bibinfo{person}{Robin Burke}, {and}
  \bibinfo{person}{Fernando Diaz}.} \bibinfo{year}{2022}\natexlab{b}.
\newblock \showarticletitle{Fairness in Recommender Systems}.
\newblock In \bibinfo{booktitle}{\emph{Recommender Systems Handbook}},
  \bibfield{editor}{\bibinfo{person}{Francesco Ricci}, \bibinfo{person}{Lior
  Rokach}, {and} \bibinfo{person}{Bracha Shapira}} (Eds.).
  \bibinfo{publisher}{Springer US}, \bibinfo{pages}{679--707}.
\newblock
\showISBNx{9781071621974}
\urldef\tempurl%
\url{https://doi.org/10.1007/978-1-0716-2197-4\_18}
\showDOI{\tempurl}


\bibitem[Ekstrand et~al\mbox{.}(2018)]%
        {ekstrand-cool-kids}
\bibfield{author}{\bibinfo{person}{Michael~D Ekstrand}, \bibinfo{person}{Mucun
  Tian}, \bibinfo{person}{Ion~Madrazo Azpiazu}, \bibinfo{person}{Jennifer~D
  Ekstrand}, \bibinfo{person}{Oghenemaro Anuyah}, \bibinfo{person}{David
  McNeill}, {and} \bibinfo{person}{Maria~Soledad Pera}.}
  \bibinfo{year}{2018}\natexlab{}.
\newblock \showarticletitle{All The Cool Kids, How Do They Fit In?: Popularity
  and Demographic Biases in Recommender Evaluation and Effectiveness}. In
  \bibinfo{booktitle}{\emph{Proceedings of the 1st Conference on Fairness,
  Accountability and Transparency}} \emph{(\bibinfo{series}{Proceedings of
  Machine Learning Research}, Vol.~\bibinfo{volume}{81})},
  \bibfield{editor}{\bibinfo{person}{Sorelle~A Friedler} {and}
  \bibinfo{person}{Christo Wilson}} (Eds.). \bibinfo{publisher}{PMLR},
  \bibinfo{pages}{172--186}.
\newblock
\urldef\tempurl%
\url{https://proceedings.mlr.press/v81/ekstrand18b.html}
\showURL{%
\tempurl}


\bibitem[Fourney et~al\mbox{.}(2018)]%
        {Fourney2018-vx}
\bibfield{author}{\bibinfo{person}{Adam Fourney}, \bibinfo{person}{Meredith
  Ringel~Morris}, \bibinfo{person}{Abdullah Ali}, {and} \bibinfo{person}{Laura
  Vonessen}.} \bibinfo{year}{2018}\natexlab{}.
\newblock \showarticletitle{Assessing the Readability of Web Search Results for
  Searchers with Dyslexia}. In \bibinfo{booktitle}{\emph{The 41st International
  {ACM} {SIGIR} Conference on Research \& Development in Information
  Retrieval}} \emph{(\bibinfo{series}{SIGIR '18})}. \bibinfo{publisher}{ACM},
  \bibinfo{pages}{1069--1072}.
\newblock
\showISBNx{9781450356572}
\urldef\tempurl%
\url{https://doi.org/10.1145/3209978.3210072}
\showDOI{\tempurl}


\bibitem[Gunawardana et~al\mbox{.}(2022)]%
        {Gunawardana2022-aw}
\bibfield{author}{\bibinfo{person}{Asela Gunawardana}, \bibinfo{person}{Guy
  Shani}, {and} \bibinfo{person}{Sivan Yogev}.}
  \bibinfo{year}{2022}\natexlab{}.
\newblock \showarticletitle{Evaluating Recommender Systems}.
\newblock In \bibinfo{booktitle}{\emph{Recommender Systems Handbook}
  (\bibinfo{edition}{third} ed.)}, \bibfield{editor}{\bibinfo{person}{Francesco
  Ricci}, \bibinfo{person}{Lior Rokach}, {and} \bibinfo{person}{Bracha
  Shapira}} (Eds.). \bibinfo{publisher}{Springer US},
  \bibinfo{pages}{547--601}.
\newblock
\showISBNx{9781071621967, 9781071621974}
\urldef\tempurl%
\url{https://doi.org/10.1007/978-1-0716-2197-4\_15}
\showDOI{\tempurl}


\bibitem[Holstein et~al\mbox{.}(2019)]%
        {Holstein2019-ho}
\bibfield{author}{\bibinfo{person}{Kenneth Holstein}, \bibinfo{person}{Jennifer
  Wortman~Vaughan}, \bibinfo{person}{Hal Daum{\'e}, III}, \bibinfo{person}{Miro
  Dudik}, {and} \bibinfo{person}{Hanna Wallach}.}
  \bibinfo{year}{2019}\natexlab{}.
\newblock \showarticletitle{Improving fairness in machine learning systems:
  What do Industry Practitioners Need?}. In
  \bibinfo{booktitle}{\emph{Proceedings of the 2019 {CHI} Conference on Human
  Factors in Computing Systems}} \emph{(\bibinfo{series}{CHI '19},
  \bibinfo{number}{Paper 600})}. \bibinfo{publisher}{ACM},
  \bibinfo{pages}{1--16}.
\newblock
\showISBNx{9781450359702}
\urldef\tempurl%
\url{https://doi.org/10.1145/3290605.3300830}
\showDOI{\tempurl}


\bibitem[Huang et~al\mbox{.}(2020)]%
        {Huang2020-na}
\bibfield{author}{\bibinfo{person}{Wen Huang}, \bibinfo{person}{Kevin Labille},
  \bibinfo{person}{Xintao Wu}, \bibinfo{person}{Dongwon Lee}, {and}
  \bibinfo{person}{Neil Heffernan}.} \bibinfo{year}{2020}\natexlab{}.
\newblock \showarticletitle{Achieving {User-Side} Fairness in Contextual
  Bandits}.
\newblock  (\bibinfo{date}{Oct.} \bibinfo{year}{2020}).
\newblock
\urldef\tempurl%
\url{http://arxiv.org/abs/2010.12102}
\showURL{%
\tempurl}


\bibitem[Huibers et~al\mbox{.}(2021)]%
        {Huibers2021-fo}
\bibfield{author}{\bibinfo{person}{Theo Huibers}, \bibinfo{person}{Monica
  Landoni}, \bibinfo{person}{Maria~Soledad Pera}, \bibinfo{person}{Jerry~Alan
  Fails}, \bibinfo{person}{Emiliana Murgia}, {and} \bibinfo{person}{Natalia
  Kucirkova}.} \bibinfo{year}{2021}\natexlab{}.
\newblock \showarticletitle{What Does Good Look Like? Report on the 3rd
  International and Interdisciplinary Perspectives on Children \& Recommender
  and Information Retrieval Systems ({KidRec}) at {IDC} 2019}.
\newblock \bibinfo{journal}{\emph{SIGIR Forum}} \bibinfo{volume}{53},
  \bibinfo{number}{2} (\bibinfo{year}{2021}), \bibinfo{pages}{76--81}.
\newblock
\showISSN{0163-5840}
\urldef\tempurl%
\url{https://doi.org/10.1145/3458553.3458561}
\showDOI{\tempurl}


\bibitem[Huibers and Westerveld(2019)]%
        {Huibers2019-el}
\bibfield{author}{\bibinfo{person}{Theo Huibers} {and} \bibinfo{person}{Thijs
  Westerveld}.} \bibinfo{year}{2019}\natexlab{}.
\newblock \showarticletitle{Relevance and utility in an educational search
  environment}. In \bibinfo{booktitle}{\emph{{KidRec} '19: Workshop in
  International and Interdisciplinary Perspectives on Children \& Recommender
  and Information Retrieval Systems, Co-located with {ACM} {IDC}}}.
\newblock
\urldef\tempurl%
\url{https://kidrec.github.io/papers/KidRec_2019_paper_4.pdf}
\showURL{%
\tempurl}


\bibitem[Kamishima et~al\mbox{.}(2018)]%
        {Kamishima2018-nn}
\bibfield{author}{\bibinfo{person}{Toshihiro Kamishima},
  \bibinfo{person}{Shotaro Akaho}, \bibinfo{person}{Hideki Asoh}, {and}
  \bibinfo{person}{Jun Sakuma}.} \bibinfo{year}{2018}\natexlab{}.
\newblock \showarticletitle{Recommendation Independence}. In
  \bibinfo{booktitle}{\emph{Proceedings of the 1st Conference on Fairness,
  Accountability and Transparency}} \emph{(\bibinfo{series}{Proceedings of
  Machine Learning Research}, Vol.~\bibinfo{volume}{81})},
  \bibfield{editor}{\bibinfo{person}{Sorelle~A Friedler} {and}
  \bibinfo{person}{Christo Wilson}} (Eds.). \bibinfo{publisher}{PMLR},
  \bibinfo{pages}{187--201}.
\newblock
\urldef\tempurl%
\url{http://proceedings.mlr.press/v81/kamishima18a.html}
\showURL{%
\tempurl}


\bibitem[Kreiss et~al\mbox{.}(2022)]%
        {Kreiss2022-lf}
\bibfield{author}{\bibinfo{person}{Elisa Kreiss}, \bibinfo{person}{Cynthia
  Bennett}, \bibinfo{person}{Shayan Hooshmand}, \bibinfo{person}{Eric
  Zelikman}, \bibinfo{person}{Meredith~Ringel Morris}, {and}
  \bibinfo{person}{Christopher Potts}.} \bibinfo{year}{2022}\natexlab{}.
\newblock \showarticletitle{Context Matters for Image Descriptions for
  Accessibility: Challenges for Referenceless Evaluation Metrics}.
\newblock  (\bibinfo{year}{2022}).
\newblock
\urldef\tempurl%
\url{http://arxiv.org/abs/2205.10646}
\showURL{%
\tempurl}


\bibitem[Li et~al\mbox{.}(2021)]%
        {Li2021-eq}
\bibfield{author}{\bibinfo{person}{Yunqi Li}, \bibinfo{person}{Hanxiong Chen},
  \bibinfo{person}{Zuohui Fu}, \bibinfo{person}{Yingqiang Ge}, {and}
  \bibinfo{person}{Yongfeng Zhang}.} \bibinfo{year}{2021}\natexlab{}.
\newblock \showarticletitle{User-oriented Fairness in Recommendation}. In
  \bibinfo{booktitle}{\emph{Proceedings of the Web Conference}}
  \emph{(\bibinfo{series}{WWW '21})}. \bibinfo{publisher}{ACM},
  \bibinfo{pages}{624--632}.
\newblock
\showISBNx{9781450383127}
\urldef\tempurl%
\url{https://doi.org/10.1145/3442381.3449866}
\showDOI{\tempurl}


\bibitem[Madras et~al\mbox{.}(2018)]%
        {Madras2018-at}
\bibfield{author}{\bibinfo{person}{David Madras}, \bibinfo{person}{Elliot
  Creager}, \bibinfo{person}{Toniann Pitassi}, {and} \bibinfo{person}{Richard
  Zemel}.} \bibinfo{year}{2018}\natexlab{}.
\newblock \showarticletitle{Learning Adversarially Fair and Transferable
  Representations}. In \bibinfo{booktitle}{\emph{Proceedings of the 35th
  International Conference on Machine Learning}}
  \emph{(\bibinfo{series}{Proceedings of Machine Learning Research},
  Vol.~\bibinfo{volume}{80})}, \bibfield{editor}{\bibinfo{person}{Jennifer Dy}
  {and} \bibinfo{person}{Andreas Krause}} (Eds.). \bibinfo{publisher}{PMLR},
  \bibinfo{pages}{3384--3393}.
\newblock
\urldef\tempurl%
\url{https://proceedings.mlr.press/v80/madras18a.html}
\showURL{%
\tempurl}


\bibitem[Madrazo~Azpiazu et~al\mbox{.}(2018)]%
        {Madrazo_Azpiazu2018-ld}
\bibfield{author}{\bibinfo{person}{Ion Madrazo~Azpiazu},
  \bibinfo{person}{Nevena Dragovic}, \bibinfo{person}{Oghenemaro Anuyah}, {and}
  \bibinfo{person}{Maria~Soledad Pera}.} \bibinfo{year}{2018}\natexlab{}.
\newblock \showarticletitle{Looking for the Movie Seven or Sven from the Movie
  Frozen? A Multi-perspective Strategy for Recommending Queries for Children}.
  In \bibinfo{booktitle}{\emph{Proceedings of the 2018 Conference on Human
  Information Interaction \& Retrieval}} \emph{(\bibinfo{series}{CHIIR '18})}.
  \bibinfo{publisher}{ACM}, \bibinfo{pages}{92--101}.
\newblock
\showISBNx{9781450349253}
\urldef\tempurl%
\url{https://doi.org/10.1145/3176349.3176379}
\showDOI{\tempurl}


\bibitem[Mauro et~al\mbox{.}(2020)]%
        {Mauro2020-be}
\bibfield{author}{\bibinfo{person}{Noemi Mauro}, \bibinfo{person}{Liliana
  Ardissono}, {and} \bibinfo{person}{Federica Cena}.}
  \bibinfo{year}{2020}\natexlab{}.
\newblock \showarticletitle{Personalized Recommendation of {PoIs} to People
  with Autism}. In \bibinfo{booktitle}{\emph{Proceedings of the 28th {ACM}
  Conference on User Modeling, Adaptation and Personalization}}
  \emph{(\bibinfo{series}{UMAP '20})}. \bibinfo{publisher}{ACM},
  \bibinfo{pages}{163--172}.
\newblock
\showISBNx{9781450368612}
\urldef\tempurl%
\url{https://doi.org/10.1145/3340631.3394845}
\showDOI{\tempurl}


\bibitem[Mehrotra et~al\mbox{.}(2017)]%
        {Mehrotra2017-ns}
\bibfield{author}{\bibinfo{person}{Rishabh Mehrotra}, \bibinfo{person}{Ashton
  Anderson}, \bibinfo{person}{Fernando Diaz}, \bibinfo{person}{Amit Sharma},
  \bibinfo{person}{Hanna Wallach}, {and} \bibinfo{person}{Emine Yilmaz}.}
  \bibinfo{year}{2017}\natexlab{}.
\newblock \showarticletitle{Auditing Search Engines for Differential
  Satisfaction Across Demographics}. In \bibinfo{booktitle}{\emph{Proceedings
  of the 26th International Conference on World Wide Web Companion}}
  \emph{(\bibinfo{series}{WWW '17 Companion})}.
  \bibinfo{publisher}{International World Wide Web Conferences Steering
  Committee}, \bibinfo{pages}{626--633}.
\newblock
\showISBNx{9781450349147}
\urldef\tempurl%
\url{https://doi.org/10.1145/3041021.3054197}
\showDOI{\tempurl}


\bibitem[Milton et~al\mbox{.}(2021)]%
        {Milton2021-rp}
\bibfield{author}{\bibinfo{person}{Ashlee Milton}, \bibinfo{person}{Garrett
  Allen}, {and} \bibinfo{person}{Maria~Soledad Pera}.}
  \bibinfo{year}{2021}\natexlab{}.
\newblock \showarticletitle{To Infinity and Beyond! Accessibility is the Future
  for Kids' Search Engines}. In \bibinfo{booktitle}{\emph{Proceedings of the
  {IR} for Children 2000-2020: Where Are We Now? Workshop co-located with
  {SIGIR} 2021}} \emph{(\bibinfo{series}{IR4C '21})}.
\newblock
\urldef\tempurl%
\url{http://arxiv.org/abs/2106.07813}
\showURL{%
\tempurl}


\bibitem[Milton et~al\mbox{.}(2020)]%
        {Milton2020-ce}
\bibfield{author}{\bibinfo{person}{Ashlee Milton}, \bibinfo{person}{Levesson
  Batista}, \bibinfo{person}{Garrett Allen}, \bibinfo{person}{Siqi Gao},
  \bibinfo{person}{Yiu-Kai~D Ng}, {and} \bibinfo{person}{Maria~Soledad Pera}.}
  \bibinfo{year}{2020}\natexlab{}.
\newblock \showarticletitle{``Don't Judge a Book by its Cover'': Exploring Book
  Traits Children Favor}. In \bibinfo{booktitle}{\emph{14th {ACM} Conference on
  Recommender Systems}} \emph{(\bibinfo{series}{RecSys '20})}.
  \bibinfo{publisher}{ACM}, \bibinfo{pages}{669--674}.
\newblock
\showISBNx{9781450375832}
\urldef\tempurl%
\url{https://doi.org/10.1145/3383313.3418490}
\showDOI{\tempurl}


\bibitem[Milton et~al\mbox{.}(2019)]%
        {Milton2019-lh}
\bibfield{author}{\bibinfo{person}{Ashlee Milton}, \bibinfo{person}{Emiliana
  Murgia}, \bibinfo{person}{Monica Landoni}, \bibinfo{person}{Theo Huibers},
  {and} \bibinfo{person}{Maria~Soledad Pera}.} \bibinfo{year}{2019}\natexlab{}.
\newblock \showarticletitle{Here, There, and Everywhere: Building a Scaffolding
  for Children's Learning through Recommendations}. In
  \bibinfo{booktitle}{\emph{Proceedings of the 1st Workshop on the Impact of
  Recommender Systems co-located with 13th {ACM} Conference on Recommender
  Systems}}, Vol.~\bibinfo{volume}{2462}. \bibinfo{publisher}{CEUR-WS}.
\newblock
\urldef\tempurl%
\url{http://ceur-ws.org/Vol-2462/short2.pdf}
\showURL{%
\tempurl}


\bibitem[Murgia et~al\mbox{.}(2019)]%
        {Murgia2019-ez}
\bibfield{author}{\bibinfo{person}{Emiliana Murgia}, \bibinfo{person}{Monica
  Landoni}, \bibinfo{person}{Theo Huibers}, \bibinfo{person}{Jerry~Alan Fails},
  {and} \bibinfo{person}{Maria~Soledad Pera}.} \bibinfo{year}{2019}\natexlab{}.
\newblock \showarticletitle{The Seven Layers of Complexity of Recommender
  Systems for Children in Educational Contexts}. In
  \bibinfo{booktitle}{\emph{Proceedings of the Workshop on Recommendation in
  Complex Scenarios co-located with 13th {ACM} Conference on Recommender
  Systems}}, Vol.~\bibinfo{volume}{2449}. \bibinfo{publisher}{CEUR-WS}.
\newblock
\urldef\tempurl%
\url{http://ceur-ws.org/Vol-2449/paper1.pdf}
\showURL{%
\tempurl}


\bibitem[Naghiaei et~al\mbox{.}(2022)]%
        {Naghiaei2022-ct}
\bibfield{author}{\bibinfo{person}{Mohammadmehdi Naghiaei},
  \bibinfo{person}{Hossein~A Rahmani}, {and} \bibinfo{person}{Yashar Deldjoo}.}
  \bibinfo{year}{2022}\natexlab{}.
\newblock \showarticletitle{{CPFair}: Personalized Consumer and Producer
  Fairness Re-ranking for Recommender Systems}. In
  \bibinfo{booktitle}{\emph{Proceedings of the 45th International {ACM} {SIGIR}
  Conference on Research and Development in Information Retrieval}}
  \emph{(\bibinfo{series}{SIGIR '22})}. \bibinfo{publisher}{ACM},
  \bibinfo{pages}{770--779}.
\newblock
\showISBNx{9781450387323}
\urldef\tempurl%
\url{https://doi.org/10.1145/3477495.3531959}
\showDOI{\tempurl}


\bibitem[Neophytou et~al\mbox{.}(2022)]%
        {Neophytou2022-cl}
\bibfield{author}{\bibinfo{person}{Nicola Neophytou}, \bibinfo{person}{Bhaskar
  Mitra}, {and} \bibinfo{person}{Catherine Stinson}.}
  \bibinfo{year}{2022}\natexlab{}.
\newblock \showarticletitle{Revisiting Popularity and Demographic Biases in
  Recommender Evaluation and Effectiveness}. In
  \bibinfo{booktitle}{\emph{Proceedings of Advances in Information Retrieval:
  44th European Conference on {IR} Research}} \emph{(\bibinfo{series}{ECIR
  '22})}. \bibinfo{publisher}{Springer-Verlag}, \bibinfo{pages}{641--654}.
\newblock
\showISBNx{9783030997359}
\urldef\tempurl%
\url{https://doi.org/10.1007/978-3-030-99736-6\_43}
\showDOI{\tempurl}


\bibitem[Netflix(2022)]%
        {netflixKids}
\bibfield{author}{\bibinfo{person}{Netflix}.} \bibinfo{year}{2022}\natexlab{}.
\newblock \bibinfo{title}{Children \& Family Movies}.
\newblock
  \bibinfo{howpublished}{\url{https://www.netflix.com/browse/genre/783}}.
\newblock


\bibitem[Ng and Pera(2018)]%
        {Ng2018-lm}
\bibfield{author}{\bibinfo{person}{Yiu-Kai Ng} {and}
  \bibinfo{person}{Maria~Soledad Pera}.} \bibinfo{year}{2018}\natexlab{}.
\newblock \showarticletitle{Recommending Social-interactive Games for Adults
  with Autism Spectrum Disorders ({ASD})}. In
  \bibinfo{booktitle}{\emph{Proceedings of the 12th {ACM} Conference on
  Recommender Systems}} \emph{(\bibinfo{series}{RecSys '18})}.
  \bibinfo{publisher}{ACM}, \bibinfo{pages}{209--213}.
\newblock
\showISBNx{9781450359016}
\urldef\tempurl%
\url{https://doi.org/10.1145/3240323.3240405}
\showDOI{\tempurl}


\bibitem[Pera et~al\mbox{.}(2019)]%
        {Pera2019-fh}
\bibfield{author}{\bibinfo{person}{Maria~Soledad Pera},
  \bibinfo{person}{Emiliana Murgia}, \bibinfo{person}{Monica Landoni}, {and}
  \bibinfo{person}{Theo Huibers}.} \bibinfo{year}{2019}\natexlab{}.
\newblock \showarticletitle{With a Little Help from My Friends: Use of
  Recommendations at School}. In \bibinfo{booktitle}{\emph{Proceedings of {ACM}
  {RecSys} 2019 Late-breaking Results}}.
\newblock
\urldef\tempurl%
\url{http://ceur-ws.org/Vol-2431/paper13.pdf}
\showURL{%
\tempurl}


\bibitem[Pera and Ng(2014)]%
        {Pera2014-ww}
\bibfield{author}{\bibinfo{person}{Maria~Soledad Pera} {and}
  \bibinfo{person}{Yiu-Kai Ng}.} \bibinfo{year}{2014}\natexlab{}.
\newblock \showarticletitle{Automating Readers' Advisory to Make Book
  Recommendations for {K-12} Readers}. In \bibinfo{booktitle}{\emph{Proceedings
  of the 8th {ACM} Conference on Recommender Systems}}
  \emph{(\bibinfo{series}{RecSys '14})}. \bibinfo{publisher}{ACM},
  \bibinfo{pages}{9--16}.
\newblock
\showISBNx{9781450326681}
\urldef\tempurl%
\url{https://doi.org/10.1145/2645710.2645721}
\showDOI{\tempurl}


\bibitem[Rastegarpanah et~al\mbox{.}(2019)]%
        {Rastegarpanah2019-qd}
\bibfield{author}{\bibinfo{person}{Bashir Rastegarpanah},
  \bibinfo{person}{Krishna~P Gummadi}, {and} \bibinfo{person}{Mark Crovella}.}
  \bibinfo{year}{2019}\natexlab{}.
\newblock \showarticletitle{Fighting Fire with Fire: Using Antidote Data to
  Improve Polarization and Fairness of Recommender Systems}. In
  \bibinfo{booktitle}{\emph{Proceedings of the 12th {ACM} International
  Conference on Web Search and Data Mining}} \emph{(\bibinfo{series}{WSDM
  '19})}. \bibinfo{publisher}{ACM}, \bibinfo{pages}{231--239}.
\newblock
\showISBNx{9781450359405}
\urldef\tempurl%
\url{https://doi.org/10.1145/3289600.3291002}
\showDOI{\tempurl}


\bibitem[Rothschild et~al\mbox{.}(2019)]%
        {Rothschild2019-xc}
\bibfield{author}{\bibinfo{person}{Meagan Rothschild}, \bibinfo{person}{Takeshi
  Horiuchi}, {and} \bibinfo{person}{Marie Maxey}.}
  \bibinfo{year}{2019}\natexlab{}.
\newblock \showarticletitle{Evaluating ``Just Right`` in {EdTech}
  Recommendation}. In \bibinfo{booktitle}{\emph{{KidRec} '19: Workshop in
  International and Interdisciplinary Perspectives on Children \& Recommender
  and Information Retrieval Systems, Co-located with {ACM} {IDC}}}.
\newblock
\urldef\tempurl%
\url{https://kidrec.github.io/papers/KidRec_2019_paper_6.pdf}
\showURL{%
\tempurl}


\bibitem[Shrestha et~al\mbox{.}(2021)]%
        {Shrestha2021-hq}
\bibfield{author}{\bibinfo{person}{Anu Shrestha}, \bibinfo{person}{Francesca
  Spezzano}, {and} \bibinfo{person}{Maria~Soledad Pera}.}
  \bibinfo{year}{2021}\natexlab{}.
\newblock \showarticletitle{An Empirical Analysis of Collaborative Recommender
  Systems Robustness to Shilling Attacks}. In
  \bibinfo{booktitle}{\emph{Proceedings of the Second Workshop on Online
  Misinformation- and {Harm-Aware} Recommender Systems co-located with 15th
  {ACM} Conference on Recommender Systems}} \emph{(\bibinfo{series}{OHARS
  '21})}. \bibinfo{publisher}{CEUR-WS}, \bibinfo{pages}{45--57}.
\newblock
\showISSN{1613-0073}
\urldef\tempurl%
\url{http://ceur-ws.org/Vol-3012/OHARS2021-paper4.pdf}
\showURL{%
\tempurl}


\bibitem[Sonboli et~al\mbox{.}(2022)]%
        {Sonboli2022-cx}
\bibfield{author}{\bibinfo{person}{Nasim Sonboli}, \bibinfo{person}{Robin
  Burke}, \bibinfo{person}{Michael Ekstrand}, {and} \bibinfo{person}{Rishabh
  Mehrotra}.} \bibinfo{year}{2022}\natexlab{}.
\newblock \showarticletitle{The Multisided Complexity of Fairness in
  Recommender Systems}.
\newblock \bibinfo{journal}{\emph{AI magazine}} \bibinfo{volume}{43},
  \bibinfo{number}{2} (\bibinfo{year}{2022}), \bibinfo{pages}{164--176}.
\newblock
\showISSN{0738-4602, 2371-9621}
\urldef\tempurl%
\url{https://doi.org/10.1002/aaai.12054}
\showDOI{\tempurl}


\bibitem[Spear et~al\mbox{.}(2021)]%
        {Spear2021-lr}
\bibfield{author}{\bibinfo{person}{Lawrence Spear}, \bibinfo{person}{Ashlee
  Milton}, \bibinfo{person}{Garrett Allen}, \bibinfo{person}{Amifa Raj},
  \bibinfo{person}{Michael Green}, \bibinfo{person}{Michael~D Ekstrand}, {and}
  \bibinfo{person}{Maria~Soledad Pera}.} \bibinfo{year}{2021}\natexlab{}.
\newblock \showarticletitle{Baby Shark to Barracuda: Analyzing Children's Music
  Listening Behavior}. In \bibinfo{booktitle}{\emph{Proceedings of the 15th
  {ACM} Conference on Recommender Systems ({RecSys} 2021 {Late-Breaking}
  Results)}}. \bibinfo{publisher}{ACM Press}.
\newblock
\urldef\tempurl%
\url{https://doi.org/10.1145/1122445.1122456}
\showDOI{\tempurl}


\bibitem[Spotify(2022)]%
        {spotifyKids}
\bibfield{author}{\bibinfo{person}{Spotify}.} \bibinfo{year}{2022}\natexlab{}.
\newblock \bibinfo{title}{Spotify Kids}.
\newblock \bibinfo{howpublished}{\url{https://www.spotify.com/us/kids/}}.
\newblock


\bibitem[Tague-Sutcliffe(1992)]%
        {Tague-Sutcliffe1992-bi}
\bibfield{author}{\bibinfo{person}{Jean Tague-Sutcliffe}.}
  \bibinfo{year}{1992}\natexlab{}.
\newblock \showarticletitle{The Pragmatics of Information Retrieval
  Experimentation, Revisited}.
\newblock \bibinfo{journal}{\emph{Information Processing \& Management}}
  \bibinfo{volume}{28}, \bibinfo{number}{4} (\bibinfo{year}{1992}),
  \bibinfo{pages}{467--490}.
\newblock
\showISSN{0306-4573}
\urldef\tempurl%
\url{https://doi.org/10.1016/0306-4573(92)90005-K}
\showDOI{\tempurl}


\bibitem[Tsiakas et~al\mbox{.}(2020)]%
        {Tsiakas2020-hm}
\bibfield{author}{\bibinfo{person}{Konstantinos Tsiakas},
  \bibinfo{person}{Emilia Barakova}, \bibinfo{person}{Javed~Vassilis Khan},
  {and} \bibinfo{person}{Panos Markopoulos}.} \bibinfo{year}{2020}\natexlab{}.
\newblock \showarticletitle{{BrainHood}: Towards an Explainable Recommendation
  System for Self-regulated Cognitive Training in Children}. In
  \bibinfo{booktitle}{\emph{Proceedings of the 13th {ACM} International
  Conference on {PErvasive} Technologies Related to Assistive Environments}}
  \emph{(\bibinfo{series}{PETRA '20}, \bibinfo{number}{Article 73})}.
  \bibinfo{publisher}{ACM}, \bibinfo{pages}{1--6}.
\newblock
\showISBNx{9781450377737}
\urldef\tempurl%
\url{https://doi.org/10.1145/3389189.3398004}
\showDOI{\tempurl}


\bibitem[Vtyurina et~al\mbox{.}(2019)]%
        {Vtyurina2019-hz}
\bibfield{author}{\bibinfo{person}{Alexandra Vtyurina}, \bibinfo{person}{Adam
  Fourney}, \bibinfo{person}{Meredith~Ringel Morris}, \bibinfo{person}{Leah
  Findlater}, {and} \bibinfo{person}{Ryen~W White}.}
  \bibinfo{year}{2019}\natexlab{}.
\newblock \showarticletitle{Bridging Screen Readers and Voice Assistants for
  Enhanced {Eyes-Free} Web Search}. In \bibinfo{booktitle}{\emph{The World Wide
  Web Conference}} \emph{(\bibinfo{series}{WWW '19})}.
  \bibinfo{publisher}{ACM}, \bibinfo{pages}{3590--3594}.
\newblock
\showISBNx{9781450366748}
\urldef\tempurl%
\url{https://doi.org/10.1145/3308558.3314136}
\showDOI{\tempurl}


\bibitem[Wang and Joachims(2021)]%
        {Wang2021-zi}
\bibfield{author}{\bibinfo{person}{Lequn Wang} {and} \bibinfo{person}{Thorsten
  Joachims}.} \bibinfo{year}{2021}\natexlab{}.
\newblock \showarticletitle{User Fairness, Item Fairness, and Diversity for
  Rankings in {Two-Sided} Markets}. In \bibinfo{booktitle}{\emph{Proceedings of
  the 2021 {ACM} {SIGIR} International Conference on Theory of Information
  Retrieval}} \emph{(\bibinfo{series}{ICTIR '21})}. \bibinfo{publisher}{ACM},
  \bibinfo{pages}{23--41}.
\newblock
\showISBNx{9781450386111}
\urldef\tempurl%
\url{https://doi.org/10.1145/3471158.3472260}
\showDOI{\tempurl}


\bibitem[Wu et~al\mbox{.}(2022)]%
        {Wu2022-fi}
\bibfield{author}{\bibinfo{person}{Haolun Wu}, \bibinfo{person}{Bhaskar Mitra},
  \bibinfo{person}{Chen Ma}, \bibinfo{person}{Fernando Diaz}, {and}
  \bibinfo{person}{Xue Liu}.} \bibinfo{year}{2022}\natexlab{}.
\newblock \showarticletitle{Joint Multisided Exposure Fairness for
  Recommendation}. In \bibinfo{booktitle}{\emph{Proceedings of the 45th
  International {ACM} {SIGIR} Conference on Research and Development in
  Information Retrieval}} \emph{(\bibinfo{series}{SIGIR '22})}.
  \bibinfo{publisher}{ACM}, \bibinfo{pages}{703--714}.
\newblock
\showISBNx{9781450387323}
\urldef\tempurl%
\url{https://doi.org/10.1145/3477495.3532007}
\showDOI{\tempurl}


\bibitem[Wu et~al\mbox{.}(2021)]%
        {Wu2021-on}
\bibfield{author}{\bibinfo{person}{Le Wu}, \bibinfo{person}{Lei Chen},
  \bibinfo{person}{Pengyang Shao}, \bibinfo{person}{Richang Hong},
  \bibinfo{person}{Xiting Wang}, {and} \bibinfo{person}{Meng Wang}.}
  \bibinfo{year}{2021}\natexlab{}.
\newblock \showarticletitle{Learning fair representations for recommendation: A
  graph-based perspective}. In \bibinfo{booktitle}{\emph{Proceedings of the Web
  Conference 2021}} \emph{(\bibinfo{series}{WebConf '21})}.
  \bibinfo{publisher}{ACM}, \bibinfo{address}{New York, NY, USA}.
\newblock
\showISBNx{9781450383127}
\urldef\tempurl%
\url{https://doi.org/10.1145/3442381.3450015}
\showDOI{\tempurl}


\bibitem[Yao and Huang(2017)]%
        {Yao2017-vz}
\bibfield{author}{\bibinfo{person}{Sirui Yao} {and} \bibinfo{person}{Bert
  Huang}.} \bibinfo{year}{2017}\natexlab{}.
\newblock \showarticletitle{Beyond Parity: Fairness Objectives for
  Collaborative Filtering}. In \bibinfo{booktitle}{\emph{Proceedings of the
  31st International Conference on Neural Information Processing Systems}}
  \emph{(\bibinfo{series}{NIPS '20})},
  \bibfield{editor}{\bibinfo{person}{I~Guyon}, \bibinfo{person}{U~V Luxburg},
  \bibinfo{person}{S~Bengio}, \bibinfo{person}{H~Wallach},
  \bibinfo{person}{R~Fergus}, \bibinfo{person}{S~Vishwanathan}, {and}
  \bibinfo{person}{R~Garnett}} (Eds.). \bibinfo{publisher}{Curran Associates,
  Inc.}, \bibinfo{pages}{2925--2934}.
\newblock
\urldef\tempurl%
\url{http://papers.nips.cc/paper/6885-beyond-parity-fairness-objectives-for-collaborative-filtering.pdf}
\showURL{%
\tempurl}


\bibitem[Zemel et~al\mbox{.}(2013)]%
        {Zemel2013-kq}
\bibfield{author}{\bibinfo{person}{Rich Zemel}, \bibinfo{person}{Yu Wu},
  \bibinfo{person}{Kevin Swersky}, \bibinfo{person}{Toni Pitassi}, {and}
  \bibinfo{person}{Cynthia Dwork}.} \bibinfo{year}{2013}\natexlab{}.
\newblock \showarticletitle{Learning Fair Representations}. In
  \bibinfo{booktitle}{\emph{Proceedings of the 30th International Conference on
  Machine Learning}} \emph{(\bibinfo{series}{Proceedings of Machine Learning
  Research}, Vol.~\bibinfo{volume}{28})},
  \bibfield{editor}{\bibinfo{person}{Sanjoy Dasgupta} {and}
  \bibinfo{person}{David McAllester}} (Eds.). \bibinfo{publisher}{PMLR},
  \bibinfo{pages}{325--333}.
\newblock
\urldef\tempurl%
\url{https://proceedings.mlr.press/v28/zemel13.html}
\showURL{%
\tempurl}


\end{thebibliography}

\end{document}